# New Perspectives and Opportunities From the Wild West of Microelectronic Biochips

Nicolò Manaresi, Gianni Medoro, Melanie Abonnenc, Vincent Auger,
Silicon Biosystems, Bologna, Italy
Paul Vulto, Aldo Romani, Luigi Altomare, Marco Tartagni, Roberto Guerrieri,
ARCES-University of Bologna, Bologna, Italy

## Abstract

*Application of Microelectronic to bioanalysis is an emerging field which holds great promise. From the standpoint of electronic and system design, biochips imply a radical change of perspective, since new, completely different constraints emerge while other usual constraints can be relaxed. While electronic parts of the system can rely on the usual established design-flow, fluidic and packaging design, calls for a new approach which relies significantly on experiments. We hereby make some general considerations based on our experience in the development of biochips for cell analysis.*

## 1. Background

The trend of miniaturization for bioanalysis is an emerging field, which is perceived to be very promising due to its potential of bringing the "cheaper, better, faster" motto to analytical practices in many application domains, such as diagnostics, biotechnology, environmental protection. Many of the early developments in this field have been based on photolithography and microfabrication of glass or plastic devices with processes borrowed from microelectronic technology. More recently, there has been a growing use of true CMOS chips for the implementation of Lab-on-a-chip [1].

In our group we pioneered the development of biochips for single-cell manipulation and detection based on CMOS [2]. An array of more than 100,000 electrodes is programmed to create electric fields in a drop of liquid (~4μl) on top of the chip, thus creating tens of thousands dielectrophoretic (DEP) cages which can trap cells in levitation [3]. Changing the pattern of voltages applied to the electrodes, the DEP cages can be shifted, thus dragging along the trapped particles. To each electrode an optical or capacitive sensor [4] can be associated, to detect particle presence.

Taking the development of these chips as a case study we can make some general considerations which apply also to other applications, and give an idea of opportunities and challenges lying ahead in this exciting brave new world of lab-on-a-chip.

## 2. Electronic design perspective: different constraints same design-flow

Fig. 1 shows a sketch of typical design flow related to integrated circuits design. High cost of prototypes, long turnaround time for fabrication, availability of accurate models favour the choice of this approach. The same is still convenient for the design of electronic part of biochips, although many constraints must be revised in light of the application. As an example, in our case, cell size is specified by biology, so there is no need to make an array with electrode pitch much smaller than that. Even more important is the fact that latest generation technologies have a reduced supply voltage while actuation (DEP force dependent on voltage square) and sensing (signal dynamic range) benefit from a larger supply voltage. This lead to a first consideration: *older generation technologies may best fit your purpose.*

Further, cells move, in response to DEP forces, at a typical rate of 10-100 microns per second, which means that we have plenty of time (from an electronic point of view) to program the actuator array, scan sensor output etc. This is an opportunity not only to achieve design goals with an older technology but also to trade time of execution for quality of the results, e.g. averaging sensors output for thermal noise reduction. Thus we can make a second consideration: *typical speeds related to transfer of mass (or heat) are quite slow compared to electronic timescale.* There is room to exploit this creatively.

## 3. Fluidic and packaging design

A very important aspect of a biochip is represented by the microfluidic and packaging implementation which represents a significant part of the design effort and of the final device cost. By fluidic, we would include here not only aspects related to the flow of liquid samples, but also, in a wider sense, the management of the sample itself for the analysis (thus including design aspects related to DEP in our example). Packaging is functional to the implementation of the desired fluidic structure, electrical connection etc. and is a key issue deeply connected with the fluidic aspects.

In this realm, although Computational Fluid Dynamics and multi-physics simulation tools are making big progresses, modeling of all relevant effect in the operation of a biochip is still a very complex task. Surface properties and wettability, heating and evaporation, electro-thermal flow, AC electro-osmosis, electric field and dielectrophoresis, modelling of cells, are all aspects requiring not only a wide knowledge of the physics behind those phenomena, but also demand -to have meaningful simulations- a lot of input parameters which are uncertain or completely unknown, thus making simulation pretty much a research topic in itself.

On the other hand, fluidic circuits can often be implemented with photolithographic processes with a moderate resolution, since minimum feature size are typically in the order of hundred microns (especially when dealing with cells which may measure 20-30μm). Moreover, fluidic design typically requires a simple mask layout (one or two layers), and fabrication can thus be achieved with a corresponding short turnaround. As a consequence *it is often faster to build and test a prototype than to simulate it*. This suggests a new approach, sketched in Fig. 2. In this perspective, simulation is not a substitute to the realization and testing of a prototype, but still has a role in helping the designer with better understanding of test results and design optimization.

In the perspective of this new design work-flow for fluidic packaging of hybrid chips we developed some special techniques [5] to achieve fast turnaround time (two-three days from design to device) and very low cost both for the masks (few euros) and overall set-up for fabrication (tens of thousands euros).

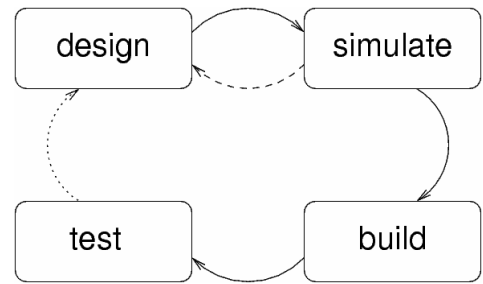

Fig. 1 - Electronic design flow. Simulation is used for verification, and further contribute to performance optimization through design centering (dashed line). After meeting specifications in simulation, one resort to fabrication and testing, with the objective to avoid lengthy and expensive further iterations (dotted line).

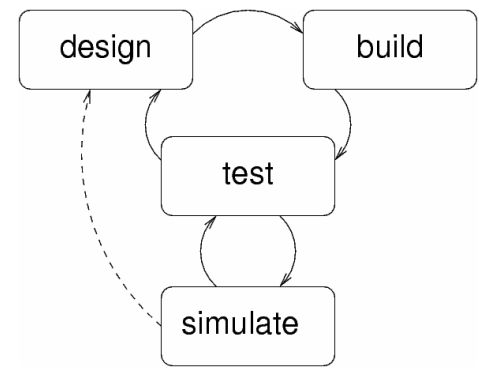

Fig. 2: Fluidic packaging design-flow. Fabrication and testing is an integral part of the design cycle. Simulation contributes with insights and interpretation of experimental data from tests, and is also useful (dashed line) to optimize the design.

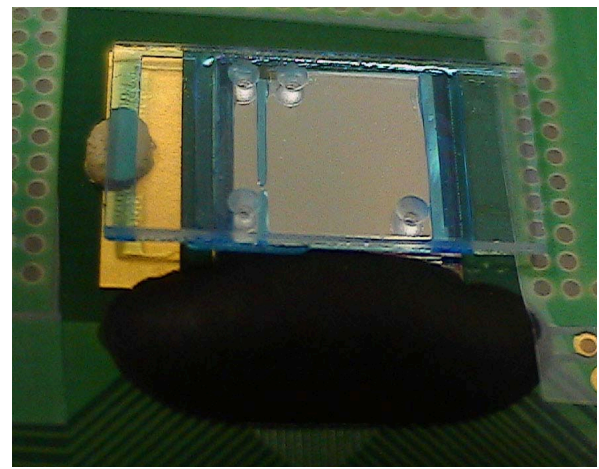

Fig. 3: CMOS biosensor/actuator for cell-analysis. The fluidic microchamber packaging is implemented double bonding the ito-coated glass, patterned with dry-resist film, to a CMOS chip.